\documentclass{article}

\usepackage{arxiv}

\usepackage[utf8]{inputenc} % allow utf-8 input
\usepackage[T1]{fontenc}    % use 8-bit T1 fonts
\usepackage[
    colorlinks=true,   % Enable colored links
    linkcolor=black,   % Color of internal links
    citecolor=black,   % Color of citations
    urlcolor=black     % Color of URLs
]{hyperref}       % hyperlinks
\usepackage{url}            % simple URL typesetting
\usepackage{booktabs}       % professional-quality tables
\usepackage{amsfonts}       % blackboard math symbols
\usepackage{nicefrac}       % compact symbols for 1/2, etc.
\usepackage{microtype}      % microtypography
\usepackage{lipsum}		% Can be removed after putting your text content
\usepackage{graphicx}
\usepackage{natbib}
\usepackage{doi}

\title{So, I climbed to the top of the pyramid of pain - now what?}

\date{30 May 2025} 					% Or removing it

\author{ \href{https://orcid.org/0000-0001-6132-3004}{\includegraphics[scale=0.06]{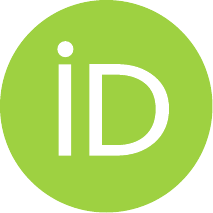}\hspace{1mm}Vasilis Katos}%\thanks{Use footnote for providing further
\\
	BU-CERT, Bournemouth University\\
	\& Cyber Innovations Ltd.\\
	\texttt{vkatos@bournemouth.ac.uk} \\
	\And
	\href{https://orcid.org/0000-0002-0481-4517}{\includegraphics[scale=0.06]{orcid.pdf}\hspace{1mm}Emily Rosenorn-Lanng} \\
	BU-CERT, Bournemouth University\\
	\& Cyber Innovations Ltd.\\
	\texttt{elanng@bournemouth.ac.uk} \\
    \And
	\href{https://orcid.org/0000-0002-0807-2661}{\includegraphics[scale=0.06]{orcid.pdf}\hspace{1mm}Jane Henriksen-Bulmer} \\
	BU-CERT, Bournemouth University\\
	\& Cyber Innovations Ltd.\\
	\texttt{jhenriksen-bulmer@bournemouth.ac.uk} \\
    \And
	\href{https://orcid.org/0000-0003-0794-0989}{\includegraphics[scale=0.06]{orcid.pdf}\hspace{1mm}Ala Yankouskaya} \\
	BU-CERT \\
	\& Department of Psychology,\\
    Bournemouth University\\
	\texttt{ayankouskaya@bournemouth.ac.uk} \\
}

\date{}

\renewcommand{\shorttitle}{Katos \textit{et al.}, 2025: So, I climbed to the top of the pyramid of pain - now what?}

\hypersetup{
pdftitle={So, I climbed to the top of the pyramid of pain - now what?},
pdfsubject={},
pdfauthor={Vasilis Katos, Emily Rosenorn-Lanng, Jane Henriksen-Bulmer, Ala Yankouskaya},
pdfkeywords={},
}

\begin{document}
\maketitle

\begin{abstract}
This paper explores the evolving dynamics of cybersecurity in the age of advanced AI, from the perspective of the introduced Human Layer Kill Chain framework. As traditional attack models like Lockheed Martin’s Cyber Kill Chain become inadequate in addressing human vulnerabilities exploited by modern adversaries, the Humal Layer Kill Chain offers a nuanced approach that integrates human psychology and behaviour into the analysis of cyber threats. We detail the eight stages of the Human Layer Kill Chain, illustrating how AI-enabled techniques can enhance psychological manipulation in attacks. By merging the Human Layer with the Cyber Kill Chain, we propose a Sociotechnical Kill Plane that allows for a holistic examination of attackers’ tactics, techniques, and procedures (TTPs) across the sociotechnical landscape. This framework not only aids cybersecurity professionals in understanding adversarial methods, but also empowers non-technical personnel to engage in threat identification and response. The implications for incident response and organizational resilience are significant, particularly as AI continues to shape the threat landscape.
\end{abstract}

% keywords can be removed

\section{Introduction}
The traditional cybersecurity landscape is evolving rapidly with the introduction of advanced AI capabilities, necessitating a reimagined approach to understanding attack methodologies. While Lockheed Martin’s Cyber Kill Chain (CKC) has provided a foundation for understanding technical attack progression, modern AI-enabled attacks increasingly target human vulnerabilities as well as technical ones. In this work we propose a Human Layer Kill Chain (HKC) framework that adapts the CKC concept to address how attackers exploit human psychology, emotions, and behaviours-especially in an era where AI technologies make traditional defensive training increasingly obsolete. We then combine the HKC with the CKC to produce the Sociotechnical Kill Chain (SKC), enabling a more granular study of the attacker’s tactics, techniques and processes (TTPs) over the whole sociotechnical landscape. Adding a second, human factors dimension on the CKC, enables not only cyber security analysts and professionals to develop a better understanding of the attacker’s TTP, but makes certain cyber threat intelligence and cyber incident response operations available to non-security and non-technical personnel.

\section{A short primer on LM's Cyber Kill Chain}
\label{sec:headings}
The Cyber Kill Chain, developed by Lockheed Martin \citep{hutchins2011intelligence}, provides a structured approach to understanding the lifecycle of cyberattacks. This model breaks down attacks into distinct phases that attackers must complete to achieve their objectives. The original CKC consists of seven phases:
\begin{enumerate}
    \item \textbf{Reconnaissance}: Gathering information about the target.
    \item \textbf{Weaponization}: Preparation of Attack Payloads.
    \item \textbf{Delivery}: The weapon is transmitted to the target environment.
    \item \textbf{Exploitation}: Triggering the attacker’s code.
    \item \textbf{Installation}: Installing malware or establishing persistence.
    \item \textbf{Command and Control}: Creating a remote manipulation channel.
    \item \textbf{Actions on Objectives}: Execution of the intended goals.
\end{enumerate}
Although the CKC has been criticised - mainly by the cyber security professionals and practitioners community - this framework has been widely adopted as a foundational model for cyber defence strategies and its added value has been acknowledged. Variations and extensions of the CKC have been proposed, such as the Unified Kill Chain \citep{pols2017unified} as well as frameworks such as MITRE’s ATT\&CK\footnote{\url{https://attack.mitre.org/}} model that curates attacker’s techniques and tactics, and the Diamond Model of Intrusion Analysis that emphasizes on the relationships of four core components of an intrusion, namely the Adversary, Infrastructure, Capability and Victim \citep{caltagirone2013diamond}.
However, with the emergence of sophisticated AI-enabled attacks that specifically target human vulnerabilities, the traditional model sems to require adaptation or extension to address the human element of security as presented in the remainder of this paper.

\section{Limitations of the existing kill chains in the AI-enabled attack era}
The traditional Cyber Kill Chain primarily focuses on technical attack vectors without adequately addressing the human factors that are increasingly exploited by modern attackers. More elaborated kill chain models such as the Unified Kill Chain (UKC) consisting of 18 stages, attempts to incorporate the human aspects by reducing them in one single stage (4th out of 18) named ``Social Engineering'', with the accompanying description of ``Techniques aimed at the psychological manipulation of people to perform unsafe actions'' \citep[p.50]{pols2017unified}. Such collapse of the description of human involvement and behaviour into one stage does not allow the capturing of the nuances, dynamics and interactions of it in critical stages of an attack, as human interaction and engagement can be required even in highly technical and automation driven attacks. Consider for example a ransomware attack. Although on UKC’s stage 4 (social engineering), user interaction \textbf{\textit{may be}} required (say, through ATT\&CK sub-technique T1566.002\footnote{\url{https://attack.mitre.org/techniques/T1566/002/}} Phishing: Spearphishing Link) in case of an attack failing to identify zero-click installation, the \textbf{\textit{“must be”}} requirement for the core objective of the attack is the victim paying the ransomware. This happens much later, during the last stage of the attack (Objectives). The coercive actions of an attacker are to instil the sense of urgency, an imminent deadline (often implemented through a countdown counter) and the option to offer a communication/chat channel in order to negotiate and generally manipulate the human operator into paying the ransom. Other, simpler attacks from a technical perspective, may require more elaborate and continuous involvement of the human. A representative example is romance scam, where the attacker must engage with the victim over a prolonged period of time, and invest on the emotional connection with them. The granularity and approaches of existing kill chains do not leave the necessary space nor provide the tools to model such interactions. 

With the rise of generative AI technologies, attacks have become more sophisticated in targeting human psychology and behaviour. In essence, several GenAI models are increasingly becoming available and able to advance capture the knowledge on human behaviour and vulnerabilities, which now can be weaponised and inform social engineering approaches \cite{kazimierczak2024impact}. In fact, we notice that a substantial amount of phishing emails over the recent year have considerably improved on the psychological manipulation attempt techniques, strongly indicating that LLMs are used as part of the attack vectors and very effectively. Moreover, generative AI systems include features that make users even more vulnerable. Their ability to simulate natural conversation, provide emotionally attuned responses, and present themselves as trustworthy companions encourages a strong sense of familiarity and reliance \citep{yankouskaya2025can}. As individuals begin to depend on these systems for emotional support, guidance, and everyday decisions, they may become less vigilant and more susceptible to influence, especially when these trusted interactions are manipulated for malicious purposes.

At the same time, traditional security awareness measures, such as phishing training, are becoming less effective as AI-enabled attacks demonstrate critical advantages such as grammatical perfection and stylistic consistency that eliminate traditional red flags; highly targeted personalization based on detailed analysis of victims; and sophisticated emotional triggers that bypass rational security awareness. As noted by \cite{da2025navigating}, “malicious actors can leverage these same tools to execute sophisticated attacks, such as highly personalized phishing schemes and the automation of malware creation” (p.1). This new reality necessitates a framework that specifically addresses how attackers target human vulnerabilities. In essence, a socio-technical system needs both a technical and human factors treatment.

\section{The human layer kill chain}
\label{HKC}
\begin{figure}[htp]
	\centering
    \includegraphics[width=0.9\textwidth]{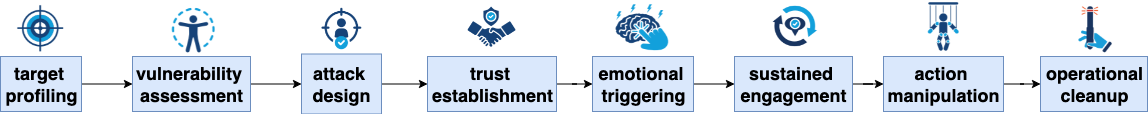}
	\caption{The Human Layer Kill Chain, HKC.}
	\label{fig:fig1}
\end{figure}

The Human Layer Kill Chain (Figure \ref{fig:fig1}) adapts the CKC concept and approach to focus specifically on how attackers exploit human psychology and behaviour, particularly in the context of AI-enhanced social engineering. This framework provides a structured approach to understanding and defending against attacks that target the human element of security.

The HKC is comprised of eight stages, with the following properties and potential AI-enabled attack capabilities:

\begin{enumerate}
    \item \textbf{Target profiling}, (e.g. OSINT/SOCMINT/HUMINT) differs from traditional reconnaissance by focusing on gathering information to create detailed psychological profiles of potential victims. In this stage, attackers seek to acquire various types of information, including personal and professional background details, emotional patterns and psychological traits, social connections and relationship dynamics, recent life events and potential stressors, as well as communication styles and preferences. AI technologies significantly enhance this phase by analysing vast amounts of data from social media profiles, professional networks, public records, and data breaches to create detailed psychological profiles. These profiles enable attackers to identify specific emotional vulnerabilities and develop highly targeted approaches.
    \item \textbf{Human vulnerability assessment.} In this stage, attackers analyse the gathered intelligence to identify specific psychological vulnerabilities in the target. This involves identifying emotional triggers that may bypass rational thinking, recognizing cognitive biases that could be exploited, and determining life circumstances and events that might render the target more susceptible. Additionally, attackers assess the target’s level of security awareness and potential blind spots, allowing them to craft more effective strategies for manipulation. AI tools can rapidly process collected intelligence to identify the most promising attack vectors based on the target’s specific psychological vulnerabilities. 
    \item \textbf{Personalized attack design.} In the personalized attack design stage, attackers utilize insights from the vulnerability assessment to craft highly personalized strategies aimed at exploiting the identified psychological vulnerabilities of their targets. This process includes creating customized narratives that resonate with the target’s specific concerns, developing content that mimics trusted entities within the target’s network, and designing emotional triggers tailored to the target’s psychological profile. Additionally, attackers prepare multiple attack paths based on anticipated responses, enhancing the likelihood of successful manipulation. Generative AI dramatically enhances this phase by enabling the creation of convincing, personalized content at scale, including deepfake videos, voice cloning, and writing that perfectly mimics trusted contacts or organizations. Such practices are already extensively found in the wild, with even state actors engaging in propaganda activities creating websites to host disinformation and propagandistic content \citep{recordedfuture2024}. These websites are not for human consumption, but for feeding distorted ground truth elements to LLMs through their data collection bots.
    \item \textbf{Trust establishment.} This critical stage involves delivering the attack through channels and contexts that the target is predisposed to trust. This process includes leveraging relationships and social proof to establish credibility, mimicking the communication patterns of trusted entities, and creating a contextual framework that makes the request appear legitimate. Additionally, attackers aim to foster a sense of familiarity by incorporating personalized details, further enhancing the likelihood of success. AI technologies enable attackers to convincingly impersonate trusted entities by analysing and replicating communication patterns, writing styles, and relationship dynamics. This makes traditional warning signs (like unfamiliar senders or generic language) obsolete as defence mechanisms.
    \item \textbf{Emotional triggering (aka ‘amygdala hijacking’).} Once trust is established, attackers activate specific emotional responses aimed at bypassing rational decision-making processes. Common emotional triggers include fear and urgency, which drive immediate action; curiosity, which can overcome caution; compassion, which encourages helping behavior; greed or opportunity, which can cloud judgment; and authority pressure, which demands compliance. AI-generated content can precisely calibrate emotional triggers based on the target’s psychological profile, making these approaches significantly more effective than traditional social engineering techniques.
    \item \textbf{Sustained engagement.} Unlike traditional attacks that may involve single-point interactions, advanced social engineering often entails ongoing engagement with the target. This stage includes maintaining believable narratives across multiple interactions, adapting approaches based on the target’s responses, and building progressive trust to enhance compliance over time. Additionally, attackers strive to create consistent experiences across different communication channels, further solidifying their manipulative efforts. AI tools enable attackers to maintain dynamic, responsive engagement by automatically generating contextually appropriate responses and adapting strategies based on the target’s reactions.
    \item \textbf{Action Manipulation.} This stage guides the target to take specific actions that align with the attacker’s objectives. These actions may include divulging sensitive information, transferring funds or assets, providing access to secured systems, making decisions that benefit the attacker, and influencing others within the organization. AI technologies help predict and shape decision-making processes by analysing patterns of behaviour and identifying the most effective manipulation techniques for each target.
    \item \textbf{Operational Cleanup.} This final stage involves covering the traces of manipulation to prevent detection and response. This may include creating alternative narratives to explain the target’s actions, employing gaslighting techniques to make the target doubt their own perceptions, and removing digital evidence of the manipulation. Additionally, attackers establish plausible deniability for themselves and work to prevent the target from recognizing that they have been manipulated, further ensuring the success of their tactics. Advanced AI techniques can help create convincing alternative explanations and remove digital traces of manipulation, making these attacks particularly difficult to detect and investigate.
\end{enumerate}

\subsection{Towards an integrated sociotechnical kill chain}
Although initiatives such as the CKC, ATT\&CK and other actionable cyber threat intelligence approaches such as MISP\footnote{\url{https://www.misp-project.org/}} touch on some concepts of human exploitation (mainly through social engineering approaches), these are superficial and primarily technology driven or described at a high level. As such, we can consider that CKC and HKC are orthogonal and can be complementary to describing attacks in a sociotechnical system. The key distinctions between these two kill chains are shown in Table \ref{tab:comp}.

\begin{table}[htp]
	\caption{CKC vs. HKC }
	\centering
\begin{tabular}{p{4cm} p{4cm} p{4cm}}
    \toprule
    \textbf{Aspect} & \textbf{Cyber Kill Chain} & \textbf{Human Layer Kill Chain} \\
    \midrule
    Primary target & Technical systems & Human psychology \\
        \addlinespace
    Attack vector & Software vulnerabilities & Emotional vulnerabilities \\
        \addlinespace
    Success factors & Technical expertise & Social engineering skills \\
        \addlinespace
    AI enhancements & Automated technical attacks & Personalised psychological manipulation \\
        \addlinespace
    Defence strategy & Technical controls, policies, processes and awareness & Psychological resilience (mental health focus) \\
    \bottomrule
\end{tabular}
	\label{tab:comp}
\end{table}

This two-dimensional enrichment can increase both the granularity and visibility when describing attack vectors and campaigns. TTP descriptions and narratives would be appropriately developed to capture, communicate, and study attacks on a sociotechnical level. The resulting 'sociotechnical kill chain' will maintain compatibility with the traditional CKC by projecting the campaign on the CKC axis or in the zero click zone. The zero-click zone serves two purposes. First, this acts as a placeholder to include phases of the attack that have not been mapped or have not been mapped yet on the HKC during an incident investigation. Second, it captures the technical phases of the attack that do not require user interaction for the underlying attack vector to progress. The \textbf{sociotechnical kill plane} is illustrated in Figure 2, and includes three representative attack vectors (romance scam, business email compromise, and ransomware) that showcase its use. 

\begin{figure}[htp]
	\centering
    \includegraphics[width=1\textwidth]{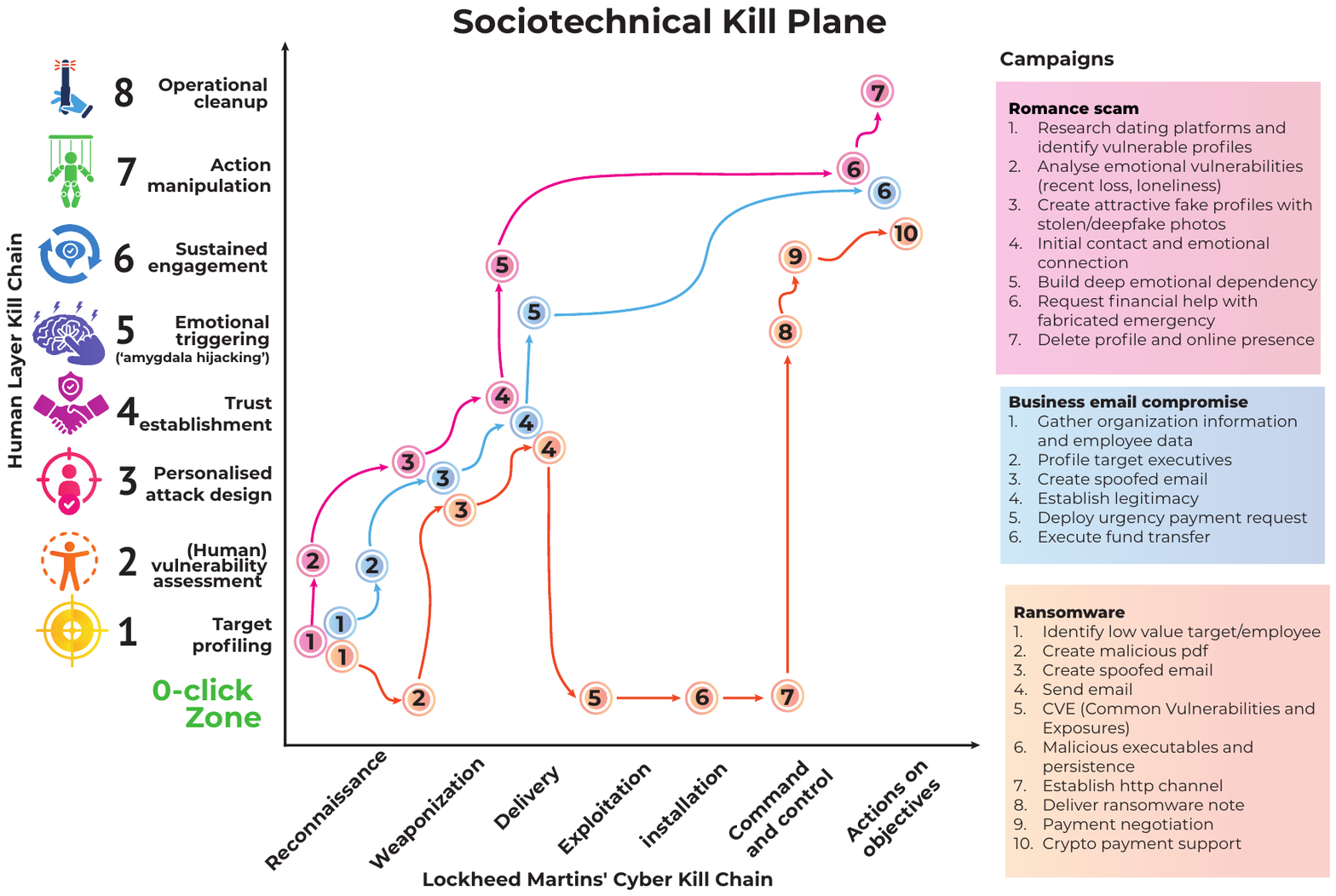}
	\caption{The integrated sociotechnical kill plane and example malicious campaigns}
	\label{fig:fig2}
\end{figure}

With the human layer kill chain being considered for analysing the attacks, more appropriate and realistic defence strategies can be developed. This particularly applies to attacks that require more substantial and meaningful user interaction – that is, attack vectors that progress outside the 0-click zone. The HKC arrangements would allow for a better understanding of such attacks when the attack descriptions are enriched with temporal information and identification of the critical HKC phase, as shown in the example in Table \ref{tab:lifecycle}. These observations may reveal the attacker’s major pain points and with the help of automated response tools (including the use of AI agents) have the potential to cause noteworthy disruption. In addition, they provide insights of suitable incident response times, with the actions of the security team staying within acceptable risk boundaries.

\begin{table}[htp]
	\caption{Attack lifecycle durations and pain points }
	\centering
\begin{tabular}{p{4cm} p{4cm} p{4cm}}
    \toprule
    \textbf{Scam type} & \textbf{Avg. duration} & \textbf{Critical HKC stage} \\
    \midrule
    Tech support & 0-48 hours	& Emotional triggering \\
        \addlinespace
    Business email compromise	& 2-14 days	& Trust establishment \\
        \addlinespace
    Romance scam	& 3-18 months	& Sustained engagement \\
    \bottomrule
\end{tabular}
	\label{tab:lifecycle}
\end{table}

\subsection{Human Indicators of Attack / Compromise (HIoAs and HIoCs)}
We define a Human Indicator of Attack (HIoA) to be an indicator that contains information that can inform a psychological tactic or technique. A Human Indicator of Compromise (HIoC) refers to observed evidence of a human operator entering a state of dysregulation following a cyber attack. HIoAs can be directly associated to traditional Indicators of Attack and can be obtained from the digital artefacts, in a similar manner as IoAs and IoCs do. In other words, IoAs are observables created and managed directly by the attacker.

\begin{figure}[htp]
	\centering
    \includegraphics[width=0.9\textwidth]{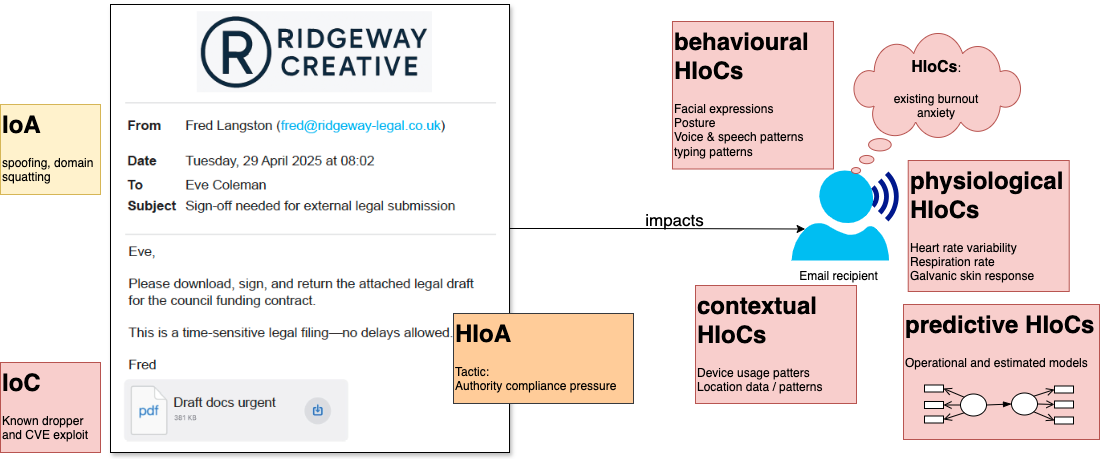}
	\caption{Indicators in a sociotechnical system}
	\label{fig:IOC}
\end{figure}

HIoCs on the other hand, refer to the psychological outcome of a human’s exposure to a HIoA. HIoCs can be directly measured through observation by third parties (e.g. team leader or line manager) or measurement technologies (generic or specialised equipment), self-reported (through scales and measures). In line with the types of IoCs – namely atomic, computed and behavioural as mentioned in LM’s CKC work – we define the following for the human layer indicator space:
\begin{itemize}
\item \textbf{Atomic HIoCs}: These are basic, individual indicators that represent directly observable behaviours or responses such as \textbf{behavioural HIoCs} (facial expressions, posture, voice and speech patterns and typing patterns) and \textbf{physiological HIoCs} (heart rate, respiration rate, galvanic skin response).
\item \textbf{Computed HIoCs}: Indicators derived from the analysis of multiple atomic indicators or data points, often resulting in complex models. These could be \textbf{contextual HIoCs} (e.g. device usage patterns, and location data/patterns) and \textbf{predictive HIoCs}, which are known, reviewed and widely acceptable statistical models capturing relations and predictor variables (such as a simple regression to integrated structural equation models). An example of a predictive HIoC would be: \textit{"Problematic internet use (i.e. digital addiction) is negatively associated with cyber security behaviour."} \citep{deutrom2022loneliness}.
\item \textbf{Latent HIoCs}: These are indicators that are not directly observable but inferred from underlying psychological states or conditions, such as depression, stress, anxiety, burnout, emotional exhaustion, and cognitive overload, etc.
\end{itemize}
The whole arrangement and distinctions of the different categories of indicators is depicted in Figure \ref{fig:IOC} using a pseudonymised example of a fictitious company.

\section{Implications}
Adding the human layer dimension on a kill chain can impact cyber incident response on a number of fronts. First, as mentioned the granularity of an attack can be studied in a much finer level. Tactics Techniques and Procedures (TTPs) can be dissected into a greater level of precision. This will not only improve cyber attribution but will also allow a better identification of attack disruption regions, which are not necessarily technical, but are located on the human layer domain. \cite{montanez2023quantifying} have initiated the creation of psychological techniques and tactics (PTechs and PTacs) loosely aligning with the Human Indicators of Attack (HIoAs) and Human Indicators of Compromise (HIoC). While the naming and typology is different from the HKC, the methodology adopted by the researchers followed the same approach for acquiring and studying malicious emails (involvement of CERTs/CSIRTs). The different findings provide some insights on the complexity and size of the problem domain and significance of attempting to study and formalise it.

At the same time, the introduction of a human dimension into what was primarily a technical capabilities focused tool (CKC), reduces the entry barrier and smoothens the learning curve to understanding cyber attacks. End users with limited expertise can focus on observing the psychological techniques and tactics of an attacker, with limited effort. This was observed during a Cyber First Aid training session, where the participants were invited to participate in a “PHINGO!” (phishing bingo) game and were asked to identify techniques and tactics over a provided set of emails. None of the participants received any prior training (other than given instructions of the bingo-based game) and all were able to participate and contribute. As such, the HKC has the potential for democratising cyber incident response, by empowering the end users to be more active participants in the identification and response of certain types of attacks.

The increased accessibility and inclusivity of the HKC, combined with the CKC in the sociotechnical kill plane could promote CKC and make it more “popular” to non-technical users. Providing the means, opportunity and option for any user or employee in general to decide on the focus of studying a threat actor’s modus operandi through an attack campaign – be it social, technical, or both – allows the individual to unfold the attack from their preferred angle and comfort zone. This will not only increase the confidence of the user but also expand and improve the cyber threat information (CTI) sharing.  

Finally, the HKC, together with the human indicators, adds temporal considerations to psychological resilience. Like estimations of brute forcing a password, further studies may provide insights into estimating the time of human burnout or mental collapse, to allow timely interventions in order to protect the users and the incident response teams.

\section{Concluding remarks}
The integration of the Human Layer Kill Chain within the broader context of cybersecurity marks a significant advancement in understanding and mitigating modern threats. As adversaries increasingly leverage AI technologies to exploit human vulnerabilities, traditional models like the Cyber Kill Chain fall short in capturing the complexities of these interactions. The HKC emphasizes the importance of human factors in cyber-attacks and provides a structured framework for analysing how psychological tactics can be weaponized.

By merging the HKC with the CKC to form the Sociotechnical Kill Chain, we create an integrated model that encompasses both technical and human dimensions of cyber threats. This dual approach enhances the granularity of threat analysis, allowing organizations to develop more effective and adaptable defence strategies. Furthermore, the HKC democratizes cybersecurity by enabling non-technical personnel to engage in threat identification and response, fostering a culture of shared responsibility for security.

Looking ahead, further research is essential to refine the HKC and explore its application across various attack scenarios. Future work should focus on developing empirical studies to identify human indicators of attack and compromise (HIoAs and HIoCs) and examining their effectiveness in real-world situations. Additionally, integrating the HKC with existing threat intelligence frameworks can enhance predictive capabilities and facilitate a more nuanced understanding of adversary behaviour. Existing kill chains and campaigns such as Unit42’s adversary playbooks\footnote{\url{https://github.com/pan-unit42/playbook_viewer/}}  can be enriched and mapped onto the two dimensions of the proposed sociotechnical kill plane, which would be particularly useful for studying different scamming techniques that have a high degree of human interaction and understanding their unique dynamics.

As we continue to navigate the evolving landscape of cyber threats, prioritizing both human vulnerabilities and technical defences will be crucial. By refining the HKC and incorporating insights from ongoing research, we can better equip ourselves to face increasingly sophisticated attacks and enhance organizational resilience.

\bibliographystyle{unsrtnat}
\bibliography{references}  %%% Uncomment this line and comment out the ``thebibliography'' section below to use the external .bib file (using bibtex) .





% \end{thebibliography}

\end{document}